\documentstyle{mn}

\input epsf

\begin{document}

\journal{astro-ph/0005085}
\title[The redshift space power spectrum in the halo model]{The redshift
space power spectrum in the halo model}
\author[M. White]{Martin White\\
Harvard-Smithsonian Center for Astrophysics\\
60 Garden St, Cambridge, MA 02138}
\date{May 2000}
\pubyear{2000}
\maketitle

\begin{abstract}
Recently there has been a lot of attention focussed on a virialized halo-based
approach to understanding the properties of the matter and galaxy power
spectrum.  We show that this model allows a natural treatment of the large and
small scale redshift space distortions, which we develop here, which extends
the pedagogical value of the approach.
\end{abstract}

\section{Introduction}

The power spectrum of the mass fluctuations in the universe is one of the
most fundamental quantities in large-scale structure.  It is robust, but
sensitive to several cosmological parameters such as the Hubble constant,
the matter density and of course the primordial power spectrum (usually
parameterized by an amplitude and a slope).
While the theory behind the power spectrum in the linear regime is quite
straightforward, analytically handling clustering in the non-linear regime
has proven quite difficult.
Recently several authors \cite{MaFry,Sel,Pea,PeaSmi} have developed a new
way of looking at the non-linear power spectrum which imagines all the mass
in the universe lies in a halo of some mass \cite{PreSch}.
They postulate that on large scales the halos cluster according to linear
theory while on small scales the power is dominated by the halo profiles
\cite{NeyScoSha,Pee}.
Specifically most pairs of dark matter particles with small interparticle
separations lie within the same halo, and thus their correlations can be
predicted by the halo profile.  While this model requires many ingredients
to be fixed by numerical experiments (typically N-body simulations)
it provides a useful structure for thinking about gravitational clustering
which gives insights into several outstanding problems
\cite{Sel,PeaSmi,SelBurPen,AtrMuc,CorHuMir}.

The work to date has all focussed on the clustering of the matter or galaxies
in ``real space'', whereas many if not most observations of clustering take
place in ``redshift space''.  It is the purpose of this work to show that
redshift space distortions can be handled naturally in the halo picture and
that doing so provides insight into some well studied phenomena.

\section{The halo model}

The model for non-linear clustering is based on the Press-Schechter
\shortcite{PreSch} theory, in which all of the mass in the universe resides
in a virialized halo of a certain mass.
The extensions to this theory introduced recently allow one to calculate the
power spectra and cross correlations between both the mass and the galaxies,
given a suitable prescription for how galaxies populate dark matter halos.
Since at present such prescriptions are somewhat {\it ad hoc\/} we shall
concentrate here on the mass power spectrum, though there is no obstacle in
principle to extending the method to galaxies.

In this model the power spectrum, $P(k)$, is the sum of two pieces.
The first is that due to a system of (smooth) halos of profile $y(k)$ laid
down with inter-halo correlations assumed to be a biased sampling of
$P_{\rm lin}(k)$.
Since the real space convolution is simply a Fourier space multiplication this
contribution is
\begin{equation}
P^{{\rm 2-halo}}(k) = P_{\rm lin}(k) \left[
  \int f(\nu) d\nu\ b(\nu) y(k; M) \right]^2
\label{eqn:twohalo}
\end{equation}
where $b(\nu)$ is the (linear) bias of a halo of mass $M(\nu)$ and $f(\nu)$
is the multiplicity function.  The peak height $\nu$ is related to the mass
of the halo through
\begin{equation}
  \nu \equiv \left( {\delta_c\over \sigma(M)} \right)^2
\end{equation}
where $\delta_c=1.69$ and $\sigma(M)$ is the rms fluctuation in the matter
density smoothed with a top-hat filter on a scale $R^3=3M/4\pi \bar{\rho}$.
Both $b$ and $f$ come from fits to N-body simulations.  We use \cite{SheTor}
\begin{equation}
  b(\nu) = 1 + {\nu-1\over\delta_c} + {2p\over\delta_c(1+\nu'^{p})}
\end{equation}
and
\begin{equation}
  \nu f(\nu) = A(1+\nu'^{-p}) \nu'^{1/2} e^{-\nu'/2}
\label{eqn:fnu}
\end{equation}
where $p=0.3$ and $\nu'=0.707\nu$.  The normalization constant $A$ is fixed
by the requirement that all of the mass lie in a given halo
\begin{equation}
  \int f(\nu) d\nu = 1 \qquad .
\end{equation}

We assume that the halos all have spherical profiles depending only on the
mass.  We neglect any substructure or halos-within-halos as this seems to be
unimportant for the power spectrum.  We take the `NFW' form \cite{NFW}
\begin{equation}
  \rho(r) = {\rho_0\over x(1+x)^2} \qquad ,
\label{eqn:nfw}
\end{equation}
where $x=r/r_s$ is a scaled radius, though our results are not particularly
sensitive to this choice \cite{Sel}.  The mass of this halo is defined to be
the virial mass $M=(4\pi/3) \delta_{\rm vir}\bar{\rho}r_{\rm vir}^3$ where
we take $\delta_{\rm vir}=200$ and define the virial radius, $r_{\rm vir}$,
as the radius within which the mean density enclosed is $\delta_{\rm vir}$
times the background density.  The concentration parameter $c=r_{\rm vir}/r_s$.
The profile can then be described by its virial mass and concentration.
N-body simulations suggest the two are related and we follow \cite{Sel} in
using
\begin{equation}
  c(M) = 10\left( {M\over M_{*}} \right)^{-0.2}
\end{equation}
where $\sigma(M_{*})=1$.

Finally then we need $y(k)$ which is the Fourier transform of
Eq.~(\ref{eqn:nfw}) normalized to unit mass.  The Fourier transform can be
done analytically
\begin{equation}
\begin{array}{ll}
  \widetilde{\rho}(k) = 4\pi\rho_0 r_s^3 \left[ \vphantom{\int} \right. &
    \cos z \left\{ {\rm Ci}([1+c]z) - {\rm Ci}(z) \right\} +  \\
&  \sin z \left\{ {\rm Si}([1+c]z) - {\rm Si}(z) \right\} - 
   \left. { {\sin cz\over (1+c)z}} \right] 
\end{array}
\end{equation}
where $z\equiv kr_s$ and the total mass is
\begin{equation}
  M = 4\pi\rho_0 r_s^3\left[ \log(1+c)-{c\over 1+c} \right]
\end{equation}

\begin{figure}
\begin{center}
\leavevmode
\epsfxsize=8cm \epsfbox{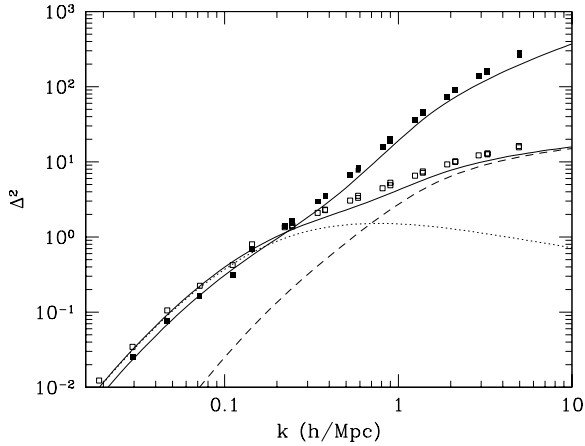}
\end{center}
\caption{The predictions of the halo model for real and redshift space
clustering, compared to a sequence of N-body simulations.
The upper solid line the sum of Eqs.~(\protect\ref{eqn:twohalo},
\protect\ref{eqn:onehalo}).  The solid squares show the results of N-body
simulations.
The lower solid line is the prediction for the redshift space power spectrum
(Eq.~\protect\ref{eqn:rs}) and the open squares the N-body results.  The
contribution of the 2-halo (dotted) and 1-halo (dashed) terms to the total
is also shown.}
\label{fig:pk}
\end{figure}

Eq.~(\ref{eqn:twohalo}) is the dominant contribution to the power spectrum
on large scales.  On small scales we are dominated by pairs lying within a
single halo
\begin{equation}
  P^{{\rm 1-halo}}(k) = {1\over (2\pi)^3} \int f(\nu) d\nu
  \ {M(\nu)\over\bar{\rho}} |y(k)|^2
\label{eqn:onehalo}
\end{equation}

We show an example of how well this formalism predicts the matter power
spectrum at $z=0$ in Fig.~\ref{fig:pk}.  We have chosen a particular
$\Lambda$CDM model with $\Omega_{\rm m}=0.3$, $\Omega_\Lambda=0.7$, $h=0.7$,
$\Omega_{\rm B}h^2=0.02$ and $n=1$.  The model has been normalized to the
{\sl COBE\/} 4-year data using the method of Bunn \& White~\shortcite{BunWhi}.
The agreement in the linear regime is good, as is to be expected.  On smaller
scales the model is a surprisingly good fit to the N-body data given the
simplicity of the assumptions.  The slight shortfall in power at higher $k$
can be remedied by modifying the prescription somewhat, but we will here stick
to the parameters outlined in \cite{Sel} --- this model is valuable more for
its pedagogical value than as a substitute for direct calculations.

\section{Redshift space distortions}

The model described above naturally lends itself to a treatment of redshift
space distortions.  There are two effects which come in when moving to
redshift space.  The first is a boost of power on large scales due to streaming
of matter into overdense regions.  The second is a reduction of power on small
scales due to virial motions within an object.  In the halo model the
large-scale and small-scale effects can be separated out in a simple fashion.  

Kaiser~\shortcite{Kai} first showed that on large scales one expects an
enhancement of the power spectrum in redshift space.
In linear theory a density perturbation $\delta_k$ generates a velocity
perturbation $\dot{\delta}=-ikv$ with $\vec{v}$ parallel to $\vec{k}$.
Using density conservation to linear order and making the distant observer
approximation ($kr\gg 1$, we shall work throughout in the plane-parallel
limit, see \cite{HeaTay,SzaMatLan} for a large-angle formalism) the redshift
space density contrast can be written as
\begin{equation}
  \delta_{\rm redshift} = \delta_{\rm real}\left( 1+ f\mu^2\right)
\end{equation}
where $f(\Omega)\equiv d\log\delta/d\log a\simeq \Omega^{0.6}$, $a$ is the
scale-factor and $\mu=\hat{r}\cdot\hat{k}$.
Thus on large-scales the power spectrum is increased by
\begin{equation}
  {1\over 2}\int_{-1}^{+1}d\mu\ \left(1+ f\mu^2\right)^2 =
  1 + {2\over 3}f + {1\over 5}f^2
\end{equation}
which is approximately $1.37$ in our model.

On small scales virial motions within collapsed objects reduce power in
redshift space.  If we assume that our halos are isotropic, virialized
and isothermal with $1D$ velocity dispersion $\sigma$ then the peculiar
motions {\it within\/} the halo add a Gaussian noise to the redshift space
radial coordinate.  Once again the real space convolution becomes a Fourier
space multiplication and the inferred density contrast is
\begin{equation}
  \delta_{\rm redshift} = \delta_{\rm real} e^{-(k\sigma\mu)^2/2}
\label{eqn:dr1dmu}
\end{equation}
Integrating this over $\mu$ gives a suppression
\begin{equation}
  {\cal R}_1(y=k\sigma) = \sqrt{{\pi\over 2}} {{\rm erf}(y/\sqrt{2})\over y}
\end{equation}

As pointed out by Peacock \& Dodds~\shortcite{PeaDodA} however, the ``full''
effect of redshift space distortions includes both the enhancement and the
suppression of power, and one must include the $\mu$ dependence of both
factors before doing the integral.
Including both terms the redshift space distortion becomes, upon integrating
over $\mu$,
\begin{eqnarray}
  {\cal R}_2(y=k\sigma) &=& {\sqrt{\pi}\over 8} { {\rm erf}(y)\over y^5}
    \left[ 3f^2+4fy^2+4y^4 \right] \nonumber \nonumber \\
  &-& {e^{-y^2}\over 4y^4}\left[ f^2(3+2y^2)+4fy^2 \right]
\end{eqnarray}

\begin{figure}
\begin{center}
\leavevmode
\epsfxsize=8cm \epsfbox{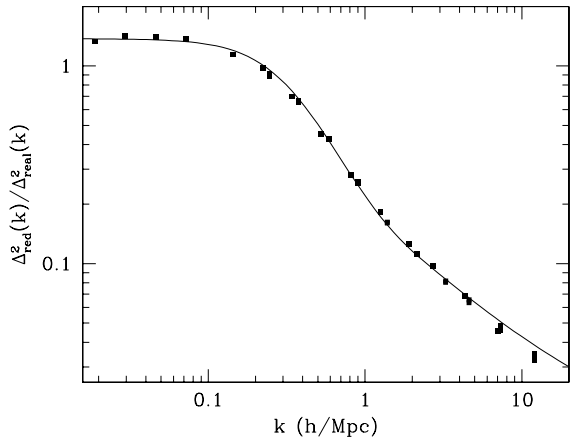}
\end{center}
\caption{The ratio of power in redshift space, compared to real space, from
the halo model (solid line) and from N-body simulations (points).}
\label{fig:ratio}
\end{figure}

Thus to predict the redshift space power spectrum in the halo model we modify
Eqs.~(\ref{eqn:twohalo}, \ref{eqn:onehalo}) to
\begin{eqnarray}
P(k) &=& \left( 1+{2\over 3}f + {1\over 5}f^2\right)P_{\rm lin}(k)\times
   \nonumber \\
& &\left[ \int f(\nu) d\nu\ b(\nu) {\cal R}_1(k\sigma) y(k; M) \right]^2
   \nonumber \\
&+& {1\over (2\pi)^3} \int f(\nu) d\nu \ {M(\nu)\over\bar{\rho}}
    {\cal R}_2(k\sigma) |y(k)|^2
\label{eqn:rs}
\end{eqnarray}
In the first term we have broken out the enhancement due to halo motions,
treated in linear theory, and the ${\cal R}_1$ term from the virial motion.
Technically we should do the integral over $\nu$ including the Gaussian from
Eq.~(\ref{eqn:dr1dmu}) first and then integrate over $\mu$.  We can see in
Fig.~\ref{fig:ratio} however that this numerically simpler approximation works
quite well.
Note that in contrast to the real-space power spectrum the clustering term
remains significant to larger $k$.

We assume that the halos are isothermal.  From the mass within the virial
radius the the 1D velocity dispersion of a halo of mass $M$ is
\begin{eqnarray}
  \sigma^2&=&GM/2r_{\rm vir} \\
          &=& G \left({\pi\over 6}M^2\bar{\rho} \delta_{\rm vir}\right)^{1/3}
\end{eqnarray}
We find that our results are almost entirely unchanged if we estimate $\sigma$
{}from the circular velocity interior to $r_s$ instead of $r_{\rm vir}$.

Putting the pieces together gives the lower solid line in Fig.~\ref{fig:pk}.
Notice that the model provides an adequate description of the redshift space
power spectrum.  The clustering term remains significant to smaller scales,
which may explain why perturbation theory results seem to work better in
redshift space than in real space and why the redshift space power spectrum
approximates the linear theory prediction over such a wide range of scales.
At small-$k$ this model makes the same predictions as linear theory with a
constant ``effective'' bias
\begin{equation}
  \left\langle b \right\rangle \equiv \int f(\nu) d\nu\ b(\nu)
\end{equation}
The high-$k$ behavior of the model qualitatively reproduces the suppression
seen in the N-body simulations.  This is not surprising since
Sheth~\shortcite{She} and Diaferio \& Geller~\shortcite{DiaGel} have shown
that a sum of Gaussian random velocities weighted by the Press-Schechter
\shortcite{PreSch} mass function provides a good description of the exponential
distribution of velocities seen in N-body simulations.

The model underestimates the N-body results in Fig.~\ref{fig:pk} in both real
and redshift space.  We show the prediction for the ratio of redshift-space to
real-space power in Fig.~\ref{fig:ratio} compared to the same ratio from the
N-body simulations.  Here we see that the agreement is very good over more
than 2 decades in length scale.

\section{Conclusions}

Many of the features of the power spectrum of density fluctuations in the
universe can be simply understood in a model based virialized halos.
Important ingredients in the model are that the halos be biased tracers of
the linear power spectrum and have a uniform profile with a correlation
between the internal structure and the mass which should span a wide range
sampling a a Press-Schechter \shortcite{PreSch} like mass function.
Within this model it is easy to account accurately for redshift space
distortions which alter the power on large and small scales.
The effects of virialized motions within halos suppress the ``Poisson'' or
1-halo term, so the redshift space power spectrum traces the linear theory
result to smaller physical scales.  This effect could be responsible for the
fact that perturbation theory is known to work better in redshift space than
real space when compared to numerical simulations.

\section*{Acknowledgments}

I thank U.~Seljak and R.~Sheth for useful comments on the manuscript.
This work was supported by a grant from the US National Science Foundation.

\end{document}